\documentstyle[12pt]{article}
\global\parskip 6pt

\normalsize

\begin{document}
%\begin{flushright}
%\hfill{}\\
%\end{flushright}
%\vspace*{1cm}
\thispagestyle{empty}
\centerline{\bf `Schwinger Model' on the Fuzzy Sphere }
\smallskip
\begin{center}
 E. Harikumar\footnote{Email: harisp@uohyd.ernet.in}\\
{{{\it School of Physics, University of Hyderabad,\\ Central University P O, Hyderabad, AP\\
India, PIN 500046}}}
\end{center}
\vskip.5cm
\begin{abstract}
{In this paper, we construct a model of spinor fields interacting with specific gauge fields on the fuzzy sphere and analyze the chiral symmetry of this `Schwinger model'. In constructing the theory of gauge fields interacting with spinors on the fuzzy sphere, we take the approach that the Dirac operator $D_q$ on the $q$-deformed fuzzy sphere $S_{qF}^2$ is the gauged Dirac operator on the fuzzy sphere. This introduces interaction between spinors and specific one parameter family of gauge fields. We also show how to express the field strength for this gauge field in terms of the Dirac operators $D_q$ and $D$ alone. Using the path integral method, we have calculated the $2n-$point functions of this model and show that, in general, they do not vanish, reflecting the chiral non-invariance of the partition function.}
\end{abstract}
\vspace*{.3cm}
\begin{center}
March 2010
\end{center}
\vspace*{.5cm}
keywords: {Non-Commutative Geometry, Quantum Groups, Lattice Quantum Field Theory}
\section{Introduction}
Fuzzy manifolds obtained by quantizing compact manifolds such as two-sphere $S^2$ are of interest as they are highly symmetric ``lattices''. Unlike the usual lattices here the symmetry of the underlying manifold is maintained. Also, the field theory models constructed on fuzzy spaces are regularized field theories with finite degrees of freedom. In recent times, the fuzzy sphere $S_{F}^2$\cite{madore} and a number of field theory models on $S_{F}^2$ are being studied extensively, both analytically as well as numerically
\cite{balbook,chu,chua,chub,sachin,sachina,sachinb,chan,chana,chanb,baez,baeza,baezb,digal,digala,martin,martina}. The study of fuzzy manifolds and noncommutative manifolds in general are of intrinsic interest from the point of noncommutative geometry\cite{connes,connesa}.

The fuzzy sphere $S_{F}^2$ is the algebra of $(N+1)\times(N+1)$-dimensional matrices. In the limit $N\to\infty$, the fuzzy sphere reduces to the commutative sphere $S^2$. Field theory models on $S_{F}^2$ and in general on any fuzzy manifolds, do exhibit many interesting features \cite{balbook}, like the absence of UV/IR mixing\cite{chu,chua,chub}, which is a characteristic feature of field theories in (non-compact) noncommutative spaces. But the fuzzy one-loop effective action do not reduce to that of the $S^2$ in the limit $N\to\infty$ where one recovers the commutative sphere $S^2$.

Using different approaches, several Dirac operators on $S_{F}^2$ were obtained and studied in \cite{grosse1,grosse2,grosse2a,watta1,watta2}. These Dirac operators are invariant under $SU(2)$ rotations which is the symmetry of underlying space $S_{F}^2.$ Generalizing the notions of spinor bundles to $S_{F}^2$, the Dirac operator on $S_{F}^2$, its spectrum and zero modes were obtained in \cite{grosse1}. Using a supersymmetric approach, this result was generalized to include interacting fermions as well as spinors of non-trivial topology\cite{grosse2,grosse2a}. It was shown that the spectrum of this Dirac operator is the same as that of the commutative sphere, but truncated (by the fuzzy cut-off). Without invoking supersymmetry, the Dirac operator on $S_{F}^2$ was constructed in\cite{watta1,watta2}. In \cite{bal1,bal1a} it was shown that the spinor theory on $S_{F}^2$ is free from the fermion doubling encountered in the usual lattice theories.

Fuzzy generalizations of other spaces like $CP^2$ were taken up in \cite{grosse3,bal2,bal2a,bal2b}. It was shown that the scalar field theory in this case is free of UV divergences. The Dirac operator was also obtained for fuzzy $CP^N$\cite{denjoe}. Symmetry aspects of noncommutative space were studied in \cite{paschke,varilly} bringing out the role of Hopf algebras. Generalization of fuzzy spaces, whose symmetries are implemented by quantum groups, were investigated in\cite{varilly,pinzul,dabrowski}. Later, generalizing the approach of \cite{grosse2,grosse2a}, a Dirac operator and chirality operator were constructed on q-deformed, fuzzy sphere\cite{us1}. Following this, the index theorem for this Dirac operator was obtained in \cite{us2}. Field theories on q-deformed fuzzy sphere were also studied in \cite{grosse3a,grosse3b}.

Construction and study of gauge theories on $S_{F}^2$ is the next logical step. Using the supersymmetric approach of \cite{grosse2,grosse2a}, differential calculus was defined on $S_{F}^2$ and the theory of gauge fields, interacting with scalar as well as spinor fields were constructed \cite{klimcik,grosse4}. An approach along the same line, but without using supersymmetry was employed in \cite{watta3}.

In this paper, we study a model of fermions interacting with (a specific class of) gauge field on the fuzzy sphere. Our approach is to treat the fermionic theory on $q$-deformed fuzzy sphere $F_{qF}^2$\cite{us1} as a theory of fermions interacting with a specific gauge field (parameterized by the deformation parameter $q$) on the fuzzy sphere. In addition, we also include the kinetic term for this gauge field, there by making it dynamical. Thus, in our framework, all the effects of $q$-deformation of the fuzzy sphere $S_{F}^2$ is taken as completely encoded in the gauge field $A_q$, defined on the fuzzy sphere $S_{F}^2$. The kinetic term, providing the dynamics for the gauge field is written in terms of the Laplacian defined on fuzzy sphere $S_{F}^2$ and that on $q$-deformed fuzzy sphere $S_{qF}^2$. Since this model is defined on the fuzzy sphere and not on $q$-deformed fuzzy sphere, operations of trace, integration etc are that of the fuzzy sphere.

In this paper, we take an approach different from those of\cite{grosse2,grosse2a,klimcik,grosse4,watta3} towards studying gauge fields on $S_{F}^2$. We consider the Dirac operator $D_q$ on the q-deformed fuzzy sphere $S_{qF}^2$\cite{us1} as the gauged Dirac operator on the fuzzy sphere. That is, we take $D_q$ as the Dirac operator with gauge field included,i.e., $D_q=D+A_q$, defined on $S_{F}^2$ where $D$ is the Dirac operator on $S_{F}^2$.  As we will see, the chirality operator on $S_{qF}^2$ is identical to that on $S_{F}^2$. Thus both the Dirac operators anti-commute with the same chirality operator. Thus, in this approach, all the effect of q-deformation of fuzzy sphere is encoded in the gauge field $A_q$. Note that $A_q$ is a specific gauge field on the fuzzy sphere $S_{F}^2$ characterized by the parameter $q$. The Laplacian on $S_{qF}^2$ (i.e., $D_{q}^2$) have q-dependent terms and a part of these are Lichnerowicz correction terms. Since we take all the $q$-dependence as coming from the gauge field $A_q$, it is natural to treat all the $q$-dependent terms in this Laplacian as the field strength associated with the gauge field $A_q$. Thus, in our approach, the Laplacian on $S_{qF}^2$ is sum of Laplacian and the field strength associated with the gauge field $A_q$ defined on $S_{F}^2$. Using this field strength associated with the specific gauge field (parameterized by $q$), we formulate the action for Schwinger model on $S_{F}^2$. We show that the action of this model is invariant under chiral transformation. Using the path integral approach, we then study the chiral anomaly. By explicitly evaluating the $2n-$point functions, we show that the non-invariance of the measure in the path integral leads to non-vanishing correlators. Here, we show the important role, the zero modes play in making the certain $2n-$point functions non-vanishing. We then show that by introducing appropriate $2k$ dependent term in the action, the measure can be made invariant under chiral transformations.

This paper is organized as follows. In the next section, we briefly recall the construction of the Dirac operator $D_q$ on q-deformed fuzzy sphere. After defining the $U_q(su(2))$-spinor bi-module, we define the chirality operator that splits the space of spinors into $\pm$-chiral subspaces. Then we obtain the Dirac operator which maps $\pm$-chiral subspaces to $\mp$-chiral subspaces. We show how to explicitly construct the eigenfunctions and eigenvalues of this Dirac operator, thereby obtaining its spectrum. In the process we also obtain the zero-modes of this Dirac operator. In section 3, we adopt the new approach of considering the Dirac operator $D_q$ on q-deformed fuzzy sphere $S_{qF}^2$ as the gauged Dirac operator on fuzzy sphere $S_{F}^2$. This gauged Dirac operator contains only a specific subclass of the gauge fields on $S_{F}^2$. This subclass of the gauge fields is characterized by the deformation parameter $q$. Thus, this gauged Dirac operator does not introduce coupling between fermions and all possible gauge fields on $S_{F}^2$, but only to a subclass of gauge fields. Then we define the actions for fermionic fields and the gauge field, respectively. In this approach, integration over the gauge fields is equivalent to integration over the deformation parameter $q$. Using the path integral methods, we also find out the general expression for correlators between the Dirac fields. In section 4, we discuss the chiral symmetry of the action. We recover the known result that the chiral anomaly arises from the measure corresponding to the zero modes of the Dirac operator alone. After analysing the contributions to anomaly from various Chern number sectors, we write down the partition function for the Schwinger model on $S_{F}^2$. This partition function breaks chiral symmetry. In section 5, we obtain the general expression for $2n$-point functions between spinor field and its conjugate for a sector with arbitrary number of zero modes. We then explicitly evaluate the 2-point functions for different sectors characterized by the Chern number. Using these results, we then obtain general expression for non-vanishing $2n$-point functions. Here, we see the dependence of $2n$-point function on the cut-off as well as the deformation parameter $q$. We then explain how a $2k$-dependent term can be included in the partition function to make chiral invariant. Our concluding remarks are given in section 6.

\section{Dirac Operator, its spectrum and zero modes on $S_{qF}^2$}
The q-deformed fuzzy sphere-$S_{qF}^2$ is the algebra ${\cal A}(N,q)$, of matrices where $q$ is the deformation parameter and $N$ denotes that the dimension of the matrices is $(N+1)\times (N+1)$. In the limit $q\to 1$, ${\cal A}(N,q)$ hence reduces to the algebra of $(N+1)\times(N+1)$ matrices. Here we briefly recall the construction of spinor module, chirality operator and Dirac operator on $S_{qF}^2$\cite{us1}. We also discuss the spectrum and zero modes of this Dirac operator. 

The symmetry algebra of the underlying space $S_{qF}^2$ is the Hopf algebra $U_q(su(2))$. Its generators $J_\pm, J_3$ satisfy the following relations:
\begin{equation}
 \left[ J_+, J_-\right]=[2J_3]_q,~~\left[J_3, J_\pm\right]=\pm J_\pm. 
\end{equation}
Here we have used the q-number, defined as
\begin{equation}
 [n]_q=\frac{q^{\frac{n}{2}}-q^{-\frac{n}{2}}}{q^{\frac{1}{2}}-q^{-\frac{1}{2}}}.
\end{equation}
The Casimir operator of the algebra is
\begin{equation}
 C=J_-J_+ +[J_3+\frac{1}{2}]_{q}^2.
\end{equation}

The Hopf algebra structure is complete only by defining the co-product, anti-pode and co-unit for $U_q(su(2))$. They are defined as
\begin{eqnarray}
 \Delta(J_\pm)&=&J_\pm\otimes K^{\frac{1}{2}}+K^{-\frac{1}{2}}\otimes J_\pm\\
\Delta(J_3)&=&I\otimes J_3 + J_3\otimes I\\
S(J_\pm)&=&-q^{\pm\frac{1}{2}}J_\pm,~S(J_3)=-J_3\\
\epsilon(I)&=&1,\epsilon(J_\pm)=0, \epsilon(J_3)=0,
\end{eqnarray}
where $K=q^{J_3}$. The spinor field on $S_{qF}^2$  is obtained by generalising the construction of the same on $S_{F}^2$\cite{grosse2,grosse2a,us1}. In this construction, one first perform ${\cal N}=1$ superextention of the Hopf fibration $S^3\to S^2$.This is derived from the mapping $C^{2,1}\to R^{3,2}$. The condition $A_{1}^\dagger A_1+A_{2}^\dagger A_2+b^\dagger b=r^2$, implements the Hopf superfibration $sS^3\to sS^2$, thus giving the spinor field on $S_{qF}^2$. This condition is a restriction on vectors of the representation space. Here, $b, b^\dagger$ are the Grassmann odd coordinated of $C^{2,1}$ and $sS^3$ is the $3$-sphere in $C^{2,1}$. 

The superfunction $\Phi$ is represented as a linear combination of monomials as
\begin{equation}
 \Phi={\cal C}_{m_1,m_2,n_1,n_2}^{\mu\nu} A_{1}^{\dagger~m_1} A_{2}^{\dagger~m_2} 
A_{1}^{n_1} A_{2}^{n_2} b^{\dagger \mu}b^\nu\label{mono}
\end{equation}
where $m_1,m_2,n_1,n_2$ are non-negative integers and $\mu,\nu=0,1$ with $\mu+\nu=1$ and
${\cal C}_{m_1,m_2,n_1,n_2}^{\mu\nu}$ are the coefficients.
In the above, $b$ and $b^\dagger$ are fermionic annihilation and creation operators. $A_\alpha$ and $A_{\alpha}^\dagger$ are  q-deformed bosonic annihilation and creation operators. They satisfy
the q-deformed oscillator algebra,
\begin{eqnarray}
& A_\alpha A_{\alpha}^\dagger-q^{\frac{1}{2}}A_{\alpha}^\dagger A_\alpha=q^{-\frac{N_\alpha}{2}}&\\
&[N_\alpha, A_{\alpha}^\dagger]=A_{\alpha}^\dagger,~~~[N_\alpha, A_{\alpha}]=-A_\alpha&.
\end{eqnarray}
The corresponding vacuum state obey $A_\alpha|0>_q=0$.

The Grassmann odd part of the superfunction $\Phi=\Phi_0(A_\alpha,A_{\alpha}^\dagger) + f(A_{\alpha}^\dagger,A_{\alpha}) b +g(A_{\alpha}^\dagger,A_{\alpha})b^\dagger +F(A_{\alpha}^\dagger,A_{\alpha})b^\dagger b$ is identified with the spinor field on $S_{qF}^2$. Thus the spinor field is
\begin{equation}
 \Psi= f(A_{\alpha}^\dagger,A_{\alpha}) b +g(A_{\alpha}^\dagger,A_{\alpha})b^\dagger.\label{spinor}
\end{equation}
It is a linear combination of monomials given in Eqn. (\ref{mono}). The Chern number of $\Psi$ is given by $2k=m_1+m_2+\mu-n_1-n_2-\nu=M-N, k\in \frac{1}{2}Z$. 

These spaces of spinor fields are denoted as $S_{k}\equiv S_{MN}$ where $M=m_1+m_2+\mu$ and $N=n_1+n_2+\nu$.
The spinors ( belonging to the spinor bi-module $S_{MN}$)  are maps from $ {\cal F}_{N}^\nu\to{\cal F}_{M}^\mu$. Here the elements of the 
space ${\cal F}_{N}^\nu$ is defined as
\begin{equation}
 |n_1, n_2;\nu>_q=\frac{1}{\sqrt{[n_1]_q![n_2]_q!}} A_{1}^{\dagger n_1}
A_{2}^{\dagger n_2}b^{\dagger\nu}|0>_q
\end{equation}
$|0>_q$ is the vacuum annihilated by the $A_\alpha$ and $b$. Thus these spinors of the bi-modules $S_{MN}\equiv S_k$ can be expressed as $(2N+1)\times (2M+1)$ matrices. 

Since the $U_q(su(2))$ has a natural action on these bi-modules, the $f$ and $g$  appearing in the spinor field in Eqn.(\ref{spinor}) can be written as the direct sum of IRRs corresponding to half-integer spins $j$ as
\begin{eqnarray}
 \frac{M}{2}\otimes\frac{N-1}{2}=|k+\frac{1}{2}|\oplus........\oplus (J-\frac{1}{2})~~~~
{\rm for~~ {\it f}}\label{irr1}\\
\frac{M-1}{2}\otimes\frac{N}{2}=|k-\frac{1}{2}|\oplus.....\oplus (J-\frac{1}{2}),~~~~~
{\rm for~~ {\it g}}\label{irr2}
\end{eqnarray}
respectively where $J=\frac{M+N}{2}$ and $2k=M-N$.

To express the spinor field as the sum of IRRs of $U_q(su(2))$, we give the expression for
the lowest and the highest states for given values of $j$ and $k$ ( with upper value of spin $J-\frac{1}{2}$ is fixed) as
\begin{eqnarray}
\Phi_{J,k,-j}^j =(A_{2}^\dagger q^{\frac{N_1}{4}})^{(j+k)}(A_1 q^{-\frac{N_2+1}{4}})^{(j-k)}\\
\Phi_{J,k,j}^j=(A_{1}^\dagger q^{-\frac{N_2}{4}})^{(j+k)}(A_2 q^{\frac{N_1+1}{4}})^{(j-k)}
\end{eqnarray}
respectively. Any other generic states $\Phi_{J,k.m}^j$ can be constructed by acting with $\bigtriangleup(J_\pm)$ on the above states.

The chirality operator $\Gamma$ is defined by its action as
\begin{equation}
 \Gamma\Psi=-[b^\dagger b,\Psi]\label{Gamma}.
\end{equation}
Here, we note that the chirality operator is defined using (undeformed) fermionic operator\footnote{Note that this chirality operator is not compatiable with the quantum group symmetry.} as on the fuzzy sphere\cite{grosse2,grosse2a} and $\Gamma^2=I$ (this will be used latter in Eqn.(\ref{fieldstrength})). Using the above chirality operator, we find that  
\begin{eqnarray}
fb=\Phi_{J,k,m}^{j+}=\Phi_{J,k+\frac{1}{2},m}^{j}b, j=|k+\frac{1}{2}|,.....,(J-\frac{1}{2})\label{phiplus}\\
gb^\dagger=\Phi_{J,k.m}^{j-}=\Phi_{J,k-\frac{1}{2},m}^{j}b^\dagger, j=|k-\frac{1}{2}|,.....,(J-\frac{1}{2})\label{phiminus}
\end{eqnarray}
are $\pm$ chiral states respectively. Thus we see that $f$ and $g$ belongs to $S_{k\pm\frac{1}{2}}$ respectively.
We also note that the above chiral states satisfy
\begin{equation}
 Tr (\Phi_{Jkm}^{j\pm}\Phi_{Jk^\prime 
m^\prime}^{j\mp})=\delta_{{k+k^\prime},0}\delta_{m+m^\prime,0}.\label{ortho}
\end{equation}
Here the $Tr$ is taken over the space $F_{N}^\nu=\frac{1}{\sqrt{[n_1]_q![n_2]_q!}}A_{n}^{\dagger~n_1}A_{2}^{\dagger~n_2}b^{\dagger~\nu}|0>_q$.

Having obtained the spinor field and the chirality operator, we now construct the Dirac operator. To this end, first a pair of operators $K_\pm$ which maps the $\pm$ chiral spaces to the $\mp$ states is constructed.
These operators are defined through their action on $\pm$-chiral states as
\begin{equation}
 K_\pm\Phi_{J,k,m}^{j\mp}\to \Phi_{J,k,m}^{j\pm}.
\end{equation}
These operators are given in terms of fermionic creation and annihilation operators and bosonic $q$-creation and $q$-annihilation operators as
\begin{eqnarray}
K_+\Phi&=&q^{-\frac{k- m}{4}} q^{-\frac{J_z}{2}} b \left [ A_{1}^\dagger \Phi
  A_{2}^\dagger q^{\frac{k}{2}}-  A_{2}^\dagger \Phi
  A_{1}^\dagger\right]b \\
K_-\Phi&=& q^{-\frac{k+ m}{4}} b^\dagger \left [ A_{1} \Phi
  A_{2} q^{\frac{k}{2}}-  A_{2}\Phi
  A_{1}\right]b^\dagger q^{-\frac{J_z}{2}}.
\end{eqnarray}
Using the above definitions, we can easily see that $K_+(gb^\dagger)\to(fb), K_-(fb)\to (gb^\dagger),
K_+(fb)=0$ and $K_-(gb^\dagger)=0$. This can be expressed as
\begin{equation}
 K_\pm\Phi_{J,k,m}^{j}=\sqrt{[j\pm k+1]_q[j\mp k]_q} ~\Phi_{J,k\pm 1,m}^{j}.
\end{equation}

Now in terms of these two operators, the Dirac operator is defined as
\begin{equation}
D_q\Psi=K_+\Phi_{Jkm}^-+K_-\Phi_{Jkm}^+\label{diact}
\end{equation}
Using this and Eqn.(\ref{phiplus}) and Eqn.(\ref{phiminus}) we find
\begin{equation}
 D_q\Phi_{J,k.m}^{j\pm}=\lambda(j,k,q)\Phi_{J,k.m}^{j\mp}\label{diphi}
\end{equation}
where 
$\lambda(j,k,q)=\sqrt{[j+\frac{1}{2}+k]_q[j+\frac{1}{2}-k]_q}=\sqrt{[j+\frac{1}{2}]_{q}^2-[k]_{q}^2}$. 
%Note that $\lambda(j,k,m)=\lambda(j,-k,q)$ since $[n]_q[m]_q=[m]_q[n]_q$. 
From Eqn.(\ref{diphi}), we can easily see that $D^2\Phi_{J,k.m}^{j\pm}=\lambda(j,k,q)^2\Phi_{J,k.m}^{j\pm}$, i.e.,  $\Phi_{J,k.m}^{j\pm}$ are eigenfunctions of $D^2$.

We also see  that the $|M-N|=2k$ zero modes of Dirac operator ( and hence that of $D_{q}^2$ also) are
\begin{eqnarray}
\Psi_{k,m,+}^{j,m_1, m_2}&=&{\cal N}A_{1}^{\dagger m_1} A_{2}^{\dagger m_2}b^\dagger,~~~~2k>0, -~{\rm chiral}\label{zmode1}\\
\Psi_{k,m,-}^{j,n_1, n_2}&=& {\cal N} A_{1}^{m_1} A_{2}^{ m_2}b,~~~~~~~~2k<0, +~{\rm chiral}\label{zmode2}
\end{eqnarray}
where ${\cal N}$ is a normalization constant. Both $\pm$ chiral zero modes have spin $j=|k|-\frac{1}{2}$.
These zero modes are normalized by a suitable choice of ${\cal N}$ and satisfy
\begin{equation}
Tr((\Psi_{k,m,\pm}^{j,m_1 m_2})^*\Psi_{k^\prime,m^\prime,\pm}^{j,m_1 m_2}=\delta_{k+k^\prime,0}\delta_{m+m^\prime,0}
\end{equation}
Here also the $Tr$ is taken over the space $F_{N}^\nu$ and we have defined 
$\Psi^*\equiv {\cal C}\Psi:= g^\star b-f^\star b^\dagger=\Gamma\Psi^\star$. Here ${\cal C}$ stands for the charge conjugation operator\cite{grosse4}.

The spinor field with a given Chern number can be now expanded in terms of the eigenvectors of $D_{q}^2$ operator as
\begin{equation}
\Psi_{2k}=\sum_{m_1,m_2} C_{m_1 m_2}^{\pm} \Psi_{\pm 0}^{m_1 m_2} +
\sum_{j=|k|+\frac{1}{2}}^{J-\frac{1}{2}} \left ( C_{km}^{j+}\Phi_{J, k,m}^{j+}+ C_{km}^{J-}\Phi_{J,k,m}^{j-}\right)\label{psi}
\end{equation}
where $C^\pm$ are Grassman coefficients.\footnote{Alternatively, we could expand the spinor field in terms of the eigenvectors of Dirac operator itself. The eigenvectors of Dirac operators are (the non-zero modes)
\begin{equation}
\Psi_{J,k,m}^{\pm}=\frac{1}{\sqrt{2}}\left[ \Phi_{J,k,m}^+ \pm \Phi_{J,k.m}^-\right]\label{foot}
\end{equation}
and their eigenvalue $\lambda(j,k)=\sqrt{[j+\frac{1}{2}+k]_q[j+\frac{1}{2}-k]_q}$.}
Having obtained the spinor field $\Psi_{2k}$, we now define the action for the fixed Chern number sector as
\begin{equation}
 S_{2k}=\frac{2\pi R^2}{[N+1]_q}Tr\left[ \Psi^*D\Psi+V(\Psi^*\Psi)\right]\label{dirac}
\end{equation}
where the $R$ appearing in the normalization factor is the radius of the sphere. In the above $\Psi^*=bg^\dagger-f^\dagger b=\Gamma\Psi^\star={\cal C}\Psi$ where ${\cal C}$ is the charge conjugation operator.

\section{Fermion Coupled to Gauge Field}
The eigenvectors of the Dirac operator on $S_{qF}^2$ are made up of the usual fermionic creation and annihilation operators and the $q$-deformed bosonic creation and annihilation operators. But the chirality operator (see Eqn.\ref{Gamma}) on $S_{qF}^2$ is made up of
fermionic operators alone and thus independent of $q$. Thus the chirality 
operators on $q$-deformed fuzzy sphere ($S_{qF}^2$) and fuzzy sphere ($S_{F}^2$) are identical. Thus we note that the Dirac operator $D$ on the fuzzy sphere and that on $q$-deformed sphere $D_q$ anti-commute with the same chirality operator $\Gamma$. 

We now consider the Dirac operator $D_q$ on the $q$-deformed fuzzy sphere as the Dirac operator on fuzzy sphere in the presence of specific gauge fields, i.e., $D_q=D + A_q$ is the gauged Dirac operator on fuzzy sphere. Thus, in this approach we have a specific gauge field $A_q$ on the fuzzy sphere, characterized by the parameter $q$ ( thus we have a one-parameter family of gauge fields.). Thus the action on fuzzy sphere for Dirac field with given Chern number $2k$ interacting with a specific gauge field is
\begin{equation}
 S_{q}^k=Tr~ \left[\Psi_{k}^* (D+A_q)\Psi_k +V(\Psi_{k}^*\Psi_k)\right]=
Tr~ \left[\Psi_{k}^* D_q\Psi_k+V(\Psi_{k}^*\Psi_k)\right]
\end{equation}
where we neglected the normalization factor $\frac{2\pi R^2}{[N+1]_q}$. The Chern number of the spinor field appearing in the above is $2k$.

The Dirac fields $\psi_k\equiv\psi_{MN}\in S_k\equiv S_{MN}$, $\psi_{k}^*\equiv(\psi_{MN})^*\in S_{-k}\equiv S_{NM}$, where $S_{MN}$ is the spinor module.
We expand the spinor field in terms of the eigenfunctions of $D_{q}^2$ (see footnote after Eqn.(\ref{psi})) as
\begin{equation} 
 \Psi_{MN}=\sum_{m_1,m_2}  
a_{m_1,m_2}^{0}\psi_{m_1,m_2}^0+\sum_{j=|k|+\frac{1}{2}}^{J-\frac{1}{2}}\sum_{m=-j}^{m=j}
\left[a_{km}^{j+} \Phi_{Jkm}^{j+} +a_{km}^{j-}\Phi_{Jkm}^{j-}\right].\label{Psiexp}
\end{equation}
Here the first term is the zero mode of $D_{q}^2$( as well as that of $D_{q}$) and the next terms are $\pm$-chiral eigenstates respectively.
Similarly we have 
\begin{equation}
(\Psi_{MN})^*=\Psi_{NM}^*=\sum_{m_1,m_2}  
{\bar b}_{m_1,m_2}^{0}\psi_{m_1,m_2}^0+\sum_{j=|k|+\frac{1}{2}}^{J-\frac{1}{2}}\sum_{m=-j}^{m=j}
\left[{\bar b}_{km}^{j+} \Phi_{Jkm}^{j+} +{\bar b}_{km}^{j-}\Phi_{Jkm}^{j-}\right].
\end{equation}
We have, using Eqn.(\ref{diphi}),
\begin{equation}
 D_q\Psi_{MN}=\sum_{j=|k|+\frac{1}{2}}^{J-\frac{1}{2}}\sum_{m=-j}^{m=j}
\left[a_{km}^{j+} \Phi_{Jkm}^{j-} +a_{km}^{j-}\Phi_{Jkm}^{j+}\right]\lambda( j,k,q).\label{dpsi}
\end{equation}
%where $\lambda(j,k,q)=\sqrt{[j+\frac{1}{2}+k]_q[j+\frac{1}{2}-k]_q}$

Using the orthogonality property  given in Eqn.(\ref{ortho}), we find that the action becomes (we set $\Psi^*\Psi=0$ from now onwards)
\begin{equation}
S_{q}^{2k}=S_{q}^{MN}=\sum_{j=|k|+\frac{1}{2}}^{J-\frac{1}{2}}\sum_{m=-j}^{m=j} 
\left( {\bar b}_{-k,-m}^{j+}a_{k,m}^{j+}+
{\bar b}_{-k,-m}^{j-}a_{k,m}^{j-}\right)\lambda(j,k,q)\label{sq}
\end{equation}
where ${\bar b}_{k,m}$ are the coefficients of expansion of $\Psi^*$.

Using these results, we can now write the partition function corresponding to spinor fields of a fixed Chern number as
\begin{eqnarray}
Z_{q}^{2k}&=&\int D(\Psi_{MN})^* D(\Psi_{MN}) e^{-S_{q}^{2k}}\nonumber\\
&=&\int \Pi_{n} d{\bar b}_{n}^k da_{n}^k e^{-S_{q}^k}.\label{z0}
\end{eqnarray}
Note here that $d{\bar b}_n$ and $da_n$ contain all allowed coefficients in the expansion of the spinor field for the given Chern number $2k$.

Using Eqns.(\ref{diact}) and (\ref{diphi}), we find
\begin{eqnarray}
D_{q}^2\Phi_{Jkm}^j&=&\frac{1}{2}(K_+K_-+K_-K_+)\Phi_{Jkm}^j\nonumber\\
&=&\frac{1}{2}([j+1+k]_q[j-k]_q+[j+1-k]_q[j+k]_q)\Phi_{Jkm}^j\nonumber\\
&=&([j+\frac{1}{2}]_{q}^2-[k-\frac{1}{2}]_{q}^2)\Phi_{Jkm}^j\label{dq2}
\end{eqnarray}
% D_q^2=\sum_{j=|k|+\frac{1}{2}}^{J-\frac{1}{2}}\sum_{m=-j}^{m=j}\left([j][j+1]-[k]^2 
%q^{\frac{2j+1}{2}}+(q^{\frac{1}{2}}-q^{-\frac{1}{2}})
%[k]^2([j]^2 q^{\frac{j+1}{2}}+[j+1]q^{\frac{j}{2}})\right)\label{dq2}
%\end{equation}
Since $D_q=D+A_q$, we obtain 
\begin{equation}
\Gamma F_{q}^{2k}(A)=D_{q}^2-D^2,\label{fieldstrength}
\end{equation}
where $F_{q}^{2k}(A)$ is the field strength for a fixed $2k$ sector and $\Gamma$ is the chirality operator. Here we take all the $q$-dependent terms in $D_{q}^2$ as the field strength associated with the gauge field $A_q$ since all the $q$-dependence of $D_q$ is through the gauge field $A_q$ alone. In the above $D^2=\sum_{j=|k|+\frac{1}{2}}^{J-\frac{1}{2}}\sum_{m=-j}^{m=j} (j(j+1)-k^2)$ is the Laplacian on $S_{F}^2$. Thus the Yang-Mills term can be written as $(D_{q}^2-D^2)^2$. This can be expressed as
\begin{eqnarray}
 S_{2k}(A)&=& \frac{2\pi R^2}{[N+1]_q}  (D_q-D)^2\nonumber\\
&=&\frac{2\pi R^2}{[N+1]_q} 
\left(\sum_{j=|k|+\frac{1}{2}}^{J-\frac{1}{2}}\sum_{m=-j}^{m=j}~(j(j+1)-k^2)\right)^2+ 
F_{q}^{2k\prime}\label{maxwell}
\end{eqnarray}
As it is clear from the notation, $F_{q}^{2k\prime}$ contains all $q$-dependent terms appearing in the field strength.

We can write down the partition function for the Schwinger model in the presence of sources ${J}_{MN}^*$ and $J_{MN}$ for ${\Psi}_{MN}^*$ and $\Psi_{MN}$ respectively as
\begin{eqnarray}
Z_{q}^{2k}[{\cal J}^*, {\cal J}]&=&\int \Pi_{n} d{\bar b}_{n}^k da_{n}^k e^{S_{q}^{2k} +S_{2k}(A)+Tr({\Psi}^*{\cal J}+{\cal J}^*\psi)}\nonumber\\
&=&e^{S_q({\cal J}^*,{\cal J})+S_{2k}(A)}\label{zj}
\end{eqnarray}
where
\begin{eqnarray}
S_q({\cal J}^*,{\cal J})&=&\sum_{j=|k|+\frac{1}{2}}^{J-\frac{1}{2}}\sum_{m=-j}^{m=j}
{\cal J}_{J,k,m}^*{\cal J}_{J,k^\prime,m^\prime} \delta_{k+k^\prime,0}\delta_{m+m^\prime,0}\frac{1}{\lambda(j,k,q)}~.\label{source}
\end{eqnarray}
In the above, we have assumed the expansion of ${{\cal J}^*}$ and ${\cal J}$ in terms of the same eigenbasis as that used to expand $\Psi_{MN}$ as in Eqn.(\ref{Psiexp}).

From this we get
\begin{equation}
 <a_{J,k,m}^{j\pm}\Phi_{J,k,m}^{j\pm},b_{J,k^\prime,m^\prime}^{j\mp}\Phi_{J,k^\prime,m^\prime}^{j\mp}>~
=~\delta_{k+k^\prime,0}\delta_{m+m^\prime,0}\frac{1}{\lambda(j,k,q)}~\label{correlator}
\end{equation}
where $\lambda(j,k,q)$ are the eigenvalues of the Dirac operator (see Eqn.(\ref{dpsi}))

\section{Chiral Transformation}
Under chiral transformation, we have
\begin{eqnarray}
 \Psi\rightarrow \Psi^\prime&=&e^{i\beta \Gamma}\Psi\nonumber\\
{\Psi}^*\rightarrow{\Psi}^{*\prime}&=& {\Psi^*}e^{i\beta\Gamma}\label{ct}
\end{eqnarray}
where $\Gamma F=-[b^\dagger b, F]$. 
Using the expansion of $\psi_{MN}$ it is easy to see that
\begin{equation}
a_{m_1 m_2}^0\to e^{i\epsilon(2k)\beta} a_{m_1 m_2}^0,~~~a_{km}^{j\pm}\to a_{km}^{j\pm} e^{\pm i\beta}
\end{equation}
where $\epsilon(2k)$ is either $\mp2k$ depending on the chirality of the zero modes ( For $2k>0$, the zero modes are $-$ chiral and for $2k<0$ they are $+$ chiral).
Using this, under the chiral transformation, we find 
\begin{equation}
[D\Psi]_{MN}\to e^{-i2k\beta}[D\Psi]_{MN}\label{ct1}
\end{equation}
\begin{equation}
[D\Psi]_{NM}^*\to e^{-i2k\beta}[D\Psi]_{NM}^*\label{ct2}
\end{equation}
It is easy to see that the actions for spinor fields (see Eqn.(\ref{dirac})) as well as that for the kinetic part of the gauge field (see Eqn.(\ref{maxwell})) are invariant under the above chiral transformation.

We observe that
\begin{enumerate}
 \item The contribution from $\pm$ chiral modes cancel each other for nonzero modes and the
non-vanishing  contributions to the measure are only from the zero modes.
\item For any given cut-off $J=\frac{M+N}{2}$, if $2k$ is an allowed Chern number, then $-2k$ is also an allowed one.
\item We have $2k_{max}=2(j+\frac{1}{2}),~2k_{min}=-2(j+\frac{1}{2})$ and $2k
=-2(j+\frac{1}{2}), -2(j+\frac{1}{2}+1),.....,2(j-\frac{1}{2}),2(j+\frac{1}{2}).$ 
\end{enumerate}
Thus we see that the measure for a sector with fixed Chern number $2k$ transforms as
\begin{equation}
D({\Psi}_{MN})^*D{\Psi}_{MN}\to e^{-4ik\beta}D({\Psi}_{MN})^*D{\Psi}_{MN}
\end{equation}
 under chiral transformation. 

Now the partition function for Schwinger model which includes sectors of spinor fields with all possible Chern numbers is 
\begin{equation}
Z_q=\sum_{k} \int \prod d{\bar b}_{n}^{k} da_{n}^{k} e^{S_{2k}+S_{2k}(A)}\label{ctnoninv01}
\end{equation}
where the summation over $k$ runs over all allowed values of $k$ ranging $K_{min}=-(j-\frac{1}{2})$ to $K_{max}=(j+\frac{1}{2})$.
Under the chiral transformation, the above partition function is not invariant as the measure picks up an exponential factor and becomes
\begin{equation}
Z_q=\sum_{k} \int \prod d{\bar b}_{n}^{k} da_{n}^{k} e^{S_{2k} +S_{2k}(A)-4i\beta k}\label{ctnoninv}
\end{equation}
Note that the $\beta$ which is the chiral transformation parameter (see Eqn.(\ref{ct})) appear in the exponent due to the chiral non-invariance of the partition function in Eqn.(\ref{ctnoninv01}).

Using Eqn.(\ref{sq}) in the above, and Eqn.(\ref{maxwell}), we find the partition function for the Schwinger model on $S_{F}^2$ for a fixed $q$ to be
\begin{eqnarray}
 Z_q
&=&\sum_{k}\left(\prod_{j=|k|+\frac{1}{2}}^{J-\frac{1}{2}}\prod_{m=-j}^{j} \lambda(j,k,q)\right)e^{S_{2k}(A)}\label{chiralnoninv}\nonumber\\
&=&\sum_{k}\left(\prod_{j=|k|+\frac{1}{2}}^{J-\frac{1}{2}} (\lambda(j,k,q))^{2j+1}\right)e^{S_{2k}(A)}.\label{ctnoninv1}
\end{eqnarray}

In the above we have used the fact that $\lambda(j,k,q)$ is the same for all allowed values of $m$ for fixed values of $j$ and $k$. Also, the summation over $k$ ranges from $k_{min}=-(j+\frac{1}{2})$ to $k_{max}=j+\frac{1}{2}$
for each $j$ value. Note that, in the above, we have now included the Yang-Mills term for the gauge field also.

By integrating over all allowed $q$, we get the partition function in the approximation of retaining only the gauge field configuration $A_q$, i.e.,
\begin{equation}
 Z=\int dq Z_q.
\end{equation}

\section{Correlators}
For any generic function $G(\Psi^*, \Psi)$, the two-point function in a given $2k$ 
sector is 
\begin{equation}
<G(\Psi^*, \Psi)>_{q}^{2k}= Z_{q}^{-1}e^{S_{2k}(A)}\int \prod da_n d{\bar b}_n e^{S_{2k}} G(\Psi_{2k}^*,\Psi_{2k})\label{2pt1}.
\end{equation}
Thus, we have
\begin{equation}
<\Psi^*, \Psi>_{q}^{2k}=Z_{q}^{-1}e^{S_{2k}(A)}\int \prod da_n d{\bar b}_n e^{S_{2k}}\sum_{i_n,j_n} a_{i_1} b_{j_1}\psi_{i_1}\psi_{j_1}^*\label{2pt}
\end{equation}
where $a_{i_1}$ and $ {b}_{j_1}$ are the Grassmannian co-efficient appearing in the expansion of $\Psi$ and $\Psi^*$ in terms of the eigenvectors $\psi_{i_1}$ and $\psi_{j_1}^*$. Since the Yang-Mills action is independent of the fermionic fields, we have pulled this term outside the integral.

From the above equation, using the properties of Grassmannian integration, we see that the two-point function of $\psi$ and $\psi^*$ identically vanishes if the number of zero modes is more than one.

The expression for the generic $2n-$point function over all allowed $2k$ sectors, when the Dirac operator have $m$ zero modes is given by
\begin{equation}
I_{2n}^{m,q}=\frac{Z_{q}^{-1}}{(n-m)!}\sum_{i_{1}^{'}..i_{(n-m)}^{'}}\left[ \sum_{i_1,..1_n;j_1....j_n}
{\cal D}(k,j,q)\epsilon_{i_1....i_n}\epsilon_{j_1...j_n}
\psi_{i_1}\psi_{j_1}^*....
\psi_{i_1}\psi_{j_n}^*\right]\label{npoint}
\end{equation}
where 
\begin{equation}
{\cal D}(k,j,q)=\sum_{k} e^{S_{2k}(A)} \frac{det^{'}D}{\lambda(j,k,q)_{i_{1}^{'}}....\lambda(j,k,q)_{i_{(n-m)}^{'}}}.
\end{equation}
Here, $i_{1}^{'},...i_{(n-m)}^{'}$  labels the non-zero modes and $\lambda(j,k,q)_{i^{'}}....\lambda(j,k,q)_{i_{(n-m)}^{'}}$ are the corresponding eigenvalues, $det^{'}D$ is the determinant of the Dirac operator after excluding the zero-modes, $i_{1},...i_{n}$ labels all modes and $\psi_{i_a}$, $\psi_{j_b}^*$ etc are the eigenfunctions of the Dirac operator (including the zero modes) for the corresponding $2k$ sectors\cite{ber}.

For a $2n$-point function, from Eqn.(\ref{2pt}) we see that the RHS will have a string of product of eigenfunctions of the form $a_{i_1}\psi_{i_1}.....a_{i_n}\psi_{i_n}
b_{j_1}\psi_{j_1}^*.....b_{j_n}\psi_{j_n}^*$. Thus if there are more than $n$ zero modes,
we will not have all of them appearing in this string of products. Hence, the rules of Grassmannian integration will set the $2n$-point function to be zero identically\cite{ber}.

Now for certain specific values of the cut-off $2J=M+N$, we explicitly evaluate the two-point functions. For $J=\frac{1}{2}$ we have $j=0$ (see Eqns.(\ref{irr1},\ref{irr2})). In this case the combinations of $M, N$ and the corresponding $2k$ values are $(0,1, -1)$ and $(1,0,1)$. For these two combinations of $M,N$, we have either annihilation operators or creation operators alone present in the spinor field. Thus these two combinations lead to zero modes (see Eqns.(\ref{zmode1},\ref{zmode2})). These zero modes $\Psi_{k,m,\pm}^{j,m_1,m_2}$ are
\begin{eqnarray}
\Psi_{1,0,+}^{0,0,0}&=&b^\dagger\\
\Psi_{-1,0,-}^{0,0,0}&=&b
\end{eqnarray}
which are $\mp$-chiral zero modes respectively. These two modes are singlets under $U_q(su(2))$.

For $J=1$, the allowed combinations $(M, N, 2k)$ are $(0,2,-2), (1,1,0)$ and $(2,0,2)$ respectively. Here again, we see that for $2k=\pm2$ we have only zero modes. For $2k=+2$, these are
\begin{eqnarray}
\Psi_{2,+\frac{1}{2},+}^{\frac{1}{2},1,0}&=&q^{-\frac{N_2}{4}} A_{1}^\dagger b^\dagger\\
\Psi_{2,-\frac{1}{2},+}^{\frac{1}{2},0,1}&=&q^{\frac{N_1}{4}} A_{2}^\dagger b^\dagger
\end{eqnarray}
and they both are $-$ chiral modes.
For $2k=-2$, both the zero modes are  $+$ chiral and they are 
\begin{eqnarray}
\Psi_{-2,-\frac{1}{2},-}^{\frac{1}{2},1,0}&=&-q^{\frac{N_2+1}{4}}A_{1} b\\
\Psi_{-2,+\frac{1}{2},-}^{\frac{1}{2},0,1}&=&q^{\frac{N_2+1}{4}}A_{2}b.
\end{eqnarray}
In contrast, for $2k=0$ (with $J=1$), we have only $\pm$ chiral non-zero modes and they are
\begin{eqnarray}
&\Phi_{1,0,+\frac{1}{2}}^{\frac{1}{2}+}=+q^{-\frac{N_2}{4}}A_{1}^\dagger b,~~~~~\Phi_{1,0,-\frac{1}{2}}^{\frac{1}{2}+}=
q^{\frac{N_1}{4}}A_{2}^\dagger b\label{pchiral}&\\
&\Phi_{1,0,+\frac{1}{2}}^{\frac{1}{2}-}=q^{\frac{N_2+1}{4}}A_2b^\dagger,~~~~~
\Phi_{1,0,-\frac{1}{2}}^{\frac{1}{2}-}=-q^{\frac{N_2+1}{4}} A_1b^\dagger\label{nchiral}&
\end{eqnarray}
respectively.

Next we calculate the 2-point function for $J=\frac{1}{2},j=0$ and $J=1,j=\frac{1}{2}$ cases. In the first case as we have seen, the allowed $2k$ values are $\pm1$. Both of these sectors have only one zero mode each. We get non-vanishing contribution for the two-point function from these $2k=\pm 1$ sectors. We find this using Eqn.(\ref{npoint}) (or Eqn.(\ref{2pt})) as
\begin{equation}
<\Psi^*\Psi>_{j=0}^q=-iZ_{q}^{-1}~2e^{S_{2k}(A)} \label{2ptj0}
\end{equation}

For $J=1,j=\frac{1}{2}$, the allowed values of $2k$ are $\pm 2$ and $0$. But as we have seen, for $2k=\pm 2$, we have only zero modes and each of these sectors have two zero modes. Thus 
the 2-point function is identically zero for these sectors. Thus the only possible non-vanishing contribution to the two-point function can come from $2k=0$ sector for the case where $J=1,j=\frac{1}{2} $. Using the Eqn.(\ref{npoint}) (or Eqn.(\ref{2pt})), we find
\begin{equation}
<\Psi^*\Psi>_{j=\frac{1}{2}}^q=0.\label{2ptjhalf}
\end{equation}
For the cut-off $J=1$, the two-point function is the sum of the contributions coming from $J=\frac{1}{2}$ and $J=1$. But note that the cut-off dependence also comes through the $Z_{q}^{-1}$ factor appearing in Eqn.(\ref{2ptj0}).

Thus we see that the two-point function does not vanish for the case where the cut-off is either  $J=\frac{1}{2}$ or $J=1$. 

The non-zero values of the two-point functions shows that the partition function, as expected, is not invariant under chiral transformations.

Next let us consider the case of $j=1, J=\frac{3}{2}$. In this case, the allowed combinations of $M,N$ and the corresponding $2k$ values are $(0,3,-3), (1,2-1), (2,1,1)$ and $(3,0,3)$. As in the previous examples, for $2k=\mp3$, we have only zero modes and they are of $\pm$ chirality, respectively. Explicitly, the $+ve$-chiral zero modes with Chern number $-3$ are $A_1A_2b, A_{1}^2b,$ and $A_{2}^2b$ and the $-ve$-chiral zero modes with Chern number $+3$ are
$A_{1}^\dagger A_{2}^\dagger b^\dagger, A_{1}^{\dagger 2}b^\dagger,$ and $A_{2}^{\dagger 2}b^\dagger$. Thus for both $2k=\pm 3$, we have three zero modes each and hence the two-point function  of $\Psi^*$ and $\Psi$ does not get any contribution from these sectors.

For $2k=\pm 1$, we have only non-zero modes. These non-zero modes
corresponding to the Chern number $+1$ are 
\begin{eqnarray}
\psi_1&=&\Phi_{J,\frac{1}{2},1}^{1+},~
\psi_2=\Phi_{J,\frac{1}{2},0}^{1+},~\psi_3=\Phi_{J,\frac{1}{2},-1}^{1+},~~~~(+{\rm Chiral})\\
\psi_4&=&\Phi_{J,\frac{1}{2},1}^{1-},~\psi_5=\Phi_{J,\frac{1}{2},0}^{1-},~\psi_6=\Phi_{J,\frac{1}{2},-1}^{1-},~~~~(-{\rm Chiral}).
\end{eqnarray}
Similarly, for the Chern number $-1$, the non-zero modes are
\begin{eqnarray}
\chi_1&=&\Phi_{J,-\frac{1}{2},-1}^{1-},~~\chi_2=\Phi_{J,-\frac{1}{2},0}^{1-},~~
\chi_3=\Phi_{J,-\frac{1}{2},1}^{1-},~~~~(-{\rm Chiral})\\
\chi_4&=&\Phi_{J,-\frac{1}{2},-1}^{1+},~~\chi_5=\Phi_{J,-\frac{1}{2},0}^{1+},~~
\chi_6=\Phi_{J,-\frac{1}{2},1}^{1+},~~~~(+{\rm Chiral})
\end{eqnarray}
We have chosen the normalization such that $Tr(\Psi_i\chi_j)=\delta_{ij}$(Note that $\psi_{i}^\dagger\equiv \chi_i$). Using the Eqn.(\ref{npoint}) (or Eqn.(\ref{2pt})), we find the contribution to 2-point function to be
\begin{equation}
 <\Psi^*,\Psi>_{j=1}^q=Z_{q}^{-1}[\psi_3\chi_6 
+\psi_2\chi_5-\chi_3\psi_6-\chi_2\psi_5-\chi_1\psi_4]=0\label{2nj1}
\end{equation}
since each of the factors appearing inside the bracket are identically zero(  which comes from Eqn.(\ref{correlator}).

It is easy to see that the only non-vanishing contributions to any 2n-point functions will come from those sectors having only zero modes. Thus for a given cut-off $J$, the allowed values of $j$ are $0,\frac{1}{2},1,....,(J-\frac{1}{2})$. For each $j$ value, the zero modes come with Chern numbers $\pm 2(j+\frac{1}{2})$ and the total number of zero modes for each $j$ value is $4|k|=4(j+\frac{1}{2}).$ With these, we re-express  Eqn.(\ref{npoint})
 for 2n-point function as
\begin{equation}
 I_{2n}^{q}=-iZ_{q}^{-1}\sum_{j=0}^{J-\frac{1}{2}}~\sum_{k=j+\frac{1}{2}}^{L(J,2n)} 
e^{S_{2k}(A)}|2k|
\label{2np}
\end{equation}
where $L(J,2n)$ means, among the allowed values of $k$, the summation over $k$ extends only up to values of $J\le 2n$ where $J=\frac{M+N}{2}$ is the fuzzy cut-off. The gauge field dependence of the result appears through $Z_{q}^{-1}$. Integrating over all allowed vales of $q$, we can get the 2n-point function for Schwinger model.

%We note that the partition function for the Schwinger model which is invariant under the %chiral transformations is obtained by adding a term (see Eqns. (\ref{ctnoninv}) and %(\ref{ctnoninv1}))
%\begin{equation}
% e^{S(\theta)}=e^{+4i\lambda k}.\label{thetaterm}
%\end{equation}
%in Eqn.(\ref{ctnoninv}). With this above term included in the partition function(after %appropriate identification of $\lambda$), we can easily see that all the 2n-point functions %will be vanishing, as expected of a model with chiral invariance. Also, it is interesting to %note that with $e^{S(\theta)}$-term included in the partition function, the $2n$-point %functions will vanish for arbitrary $q$.

\section{Conclusion}

As stated in Eqn.(\ref{foot}), we could have expanded the spinor field in terms of the eigenfunctions of the Dirac operator. But note that the zero modes used in Eqn.(\ref{psi}) are that of Dirac operator itself (as they would be any way, zero modes of $D_{q}^2$). This change would not have altered our final results. As we have seen, the chiral symmetry is broken by the measure corresponding to the zero modes alone. Of course, the explicit form of the action appearing in the partition function would have been different. But even in this case, one can easily show that the end results of $2n$-point functions would be the same, by repeating exactly the same calculations we have done.

The interpretation of $D_q$ as the gauged Dirac operator $D+A_q$ on $S_{F}^2$ was crucial for treating the spinor field as coupled to the gauge field $A_q$. But equally important is that it is this which allowed us to define the field strength (see Eqn.(\ref{maxwell})) corresponding to the gauge field $A_q$. This was done using the fact that the square of the Dirac operator is nothing but the sum of Laplacian and the field strength (see Eqn.(\ref{dq2})). And it is this that leads to the introduction of kinetic term for the gauge field and thereby enabled us to define the action for Schwinger model.

We have also seen that, for any/every given value of $j$, the sector with Chern number $|2k|=2(j+\frac{1}{2})$ have only zero modes and for every zero mode with positive Chern number, we had another one with negative Chern number. This is important for the $2n$-point functions to vanish, after introducing the $S(\theta)$-term. Also, equally interesting was the observation that the other allowed Chern sectors have only non-zero modes and the contribution to $2n$-point functions from these sectors are identically vanishing. For, if this was not the case, even after introducing the $e^{S(\theta)}$-term, the $2n$-point functions would have been non-zero.

We see from Eqn.(\ref{2np}) that the $2n$-point function depends on $q$ thorough 
\begin{enumerate}
 \item $Z_{q}^{-1}$, which is a common factor and thus have same contribution for all 
allowed Chern sectors.
\item $e^{S_{2k}(A)}$ which is different for different Chern sectors.
\end{enumerate}
The cut-off dependence of the $2n$-point function come through both $Z_{q}^{-1}$ and allowed values of $2k$ appearing in the summation in Eqn. (\ref{2np}).

We can add a mass term 
\begin{eqnarray}
m^2 Tr{\bar\Psi}\Psi&=&m^2\sum_{{m_1},{m_2}}{\bar b}_{{m_1},{m_2}}^0 a_{{m_1},{m_2}}^{0}\nonumber\\ &+&m^2\sum_{j=|k|+\frac{1}{2}}^{J-\frac{1}{2}}\sum_{m=-j}^{m=j}
\left({\bar b}_{-k,-m}^{j+}a_{k,m}^{j-}+{\bar b}_{-k,-m}^{j-}a_{k,m}^{j+}\right)
\end{eqnarray}
 to the fermionic action, which will break the chiral invariance. Here, the first 
term on LHS is the contribution from zero modes. There is no such analogues terms in the massless action. Thus, it is easy to repeat the calculations of  Eqn.(\ref{source}) and Eqn.(\ref{correlator}) (correlators using the source). The only modification of this correlator would be a shift in $\lambda(j,k,q)$ by  $m^2$. Similarly, calculations of $2n$-point function can be easily done with the mass term, but the final result will be different. For example, calculation leading to Eqn.(\ref{2nj1}) would give $m^2\sum_i \chi_i\psi_i\ne 0$. Thus only in the limit $m\to 0$, $2n$-point functions will vanish, as expected of a theory without chiral symmetry.

We see that the field strength (Eqn.(\ref{fieldstrength})) is essentially expressed 
in terms of $q$-numbers and for $q\ne 1$, it is not possible to express the field strength in terms of usual numbers. In other words, the field strength will be given in terms of $q$-numbers only. Since $\beta k$ appearing in the  Eqn.(\ref{ctnoninv}) is not $q$-number, it can not be expressed in terms of the field strength.

\vspace{1cm}
\noindent{\bf ACKNOWLEDGMENTS}\\
EH would like to thank A. P. Balachandran for suggesting this problem, useful discussions and comments. He also thanks B. P. Dolan for useful discussions.

\end{document}